\documentclass[a4paper,11pt]{article}

\usepackage{jinstpub}
\usepackage{graphicx}
\usepackage{xcolor}
\usepackage{caption}
\usepackage{subcaption}
\usepackage{multirow}
\usepackage[numbers,sort&compress]{natbib}
\usepackage{verbatim}
\usepackage{textgreek}

\usepackage[utf8]{inputenc}

\title{JUNO 20-inch PMT and electronics system characterization using large pulses of PMT dark counts at the Pan-Asia testing platform}

\author[a,b]{Caimei Liu,}
\author[a,b]{Min Li,}
\author[a,b]{Narongkiat Rodphai,}
\author[a,b,1]{Zhimin Wang,\note{Corresponding author.}}
\author[a]{Jun Hu,}
\author[f]{Nikolay Anfimov,}
\author[a]{Lei Fan,}
\author[g]{Alberto Garfagnini,}
\author[h]{Guanghua Gong,}
\author[a]{Shaojing Hou,}
\author[a]{Xiaolu Ji,}
\author[a]{Xiaoshan Jiang,}
\author[f]{Denis Korablev,}
\author[c]{Tobias Lachenmaier,}
\author[a]{Si Ma,}
\author[a]{Xiaoyan Ma,}
\author[a]{Zhe Ning,}
\author[f]{Alexander G. Olshevskiy,}
\author[a,b]{Zhaoyuan Peng,}
\author[a]{Zhonghua Qin,}
\author[c]{Tobias Sterr,}
\author[a]{Yunhua Sun,}
\author[c]{Alexander Felix Tietzsch,}
\author[e]{Jun Wang,}
\author[e]{Wei Wang,}
\author[a,b]{Yangfu Wang,}
\author[a]{Kaile Wen,}
\author[d]{Bjoern Soenke Wonsak,}
\author[a]{Wan Xie,}
\author[a]{Meihang Xu,}
\author[a,b]{Xiongbo Yan,}
\author[i]{Yifan Yang,}
\author[e]{Rong Zhao,}
\author[a,b]{Tong Zhou,}
\author[a]{Kejun Zhu,}
\author[h]{Jianmeng Dong,}
\author[i]{Pierre-Alexandre PETITJEAN,}
\author[i]{Barbara CLERBAUX}

\affiliation[a]{Institute of High Energy Physics, Beijing 100049, China}
\affiliation[b]{University of Chinese Academy of Sciences, Beijing 100049, China}
\affiliation[c]{Eberhard Karls Universität Tübingen, Physikalisches Institut, Tübingen, Germany}
\affiliation[d]{Institute of Experimental Physics, University of Hamburg, Hamburg, Germany}
\affiliation[e]{Sun Yat-Sen University, Guangzhou, China}
\affiliation[f]{Joint Institute for Nuclear Research, Dubna, Russia}
\affiliation[g]{Dipartimento di Fisica e Astronomia dell'Universita' di Padova and INFN Sezione di Padova, Padova, Italy}
\affiliation[h]{Tsinghua University, Beijing, China}
\affiliation[i]{Université Libre de Bruxelles, Brussels, Belgium}
\emailAdd{wangzhm@ihep.ac.cn}

\abstract{ 
The main goal of the JUNO experiment is to determine the neutrino mass ordering with a 20\,kt liquid-scintillator detector. The 20-inch PMT and its 1F3 (one for three) electronics are crucial to realize the excellent energy resolution of at least 3\% at 1\,MeV. The knowledge on the PMT and 1F3 electronics response is critical for detector performance understanding. A study of the JUNO 20-inch PMT and 1F3 electronics system characterization is presented using large pulses of PMT dark count at the Pan-Asia testing platform in China. Thanks to its broad amplitude range and high rate, the large pulse signals are also used to investigate the PMT after pulse response.
}

\keywords{photon detectors for UV, visible and IR photons (vacuum) (photomultipliers, HPDs, others), JUNO, PMT large pulses, after pulse, JUNO 1F3 electronics}

\begin{document}
\maketitle   
\flushbottom

\section{Introduction}
\label{1:intro}

\begin{figure}[!htb]
    \centering
    \includegraphics[width=0.75\linewidth]{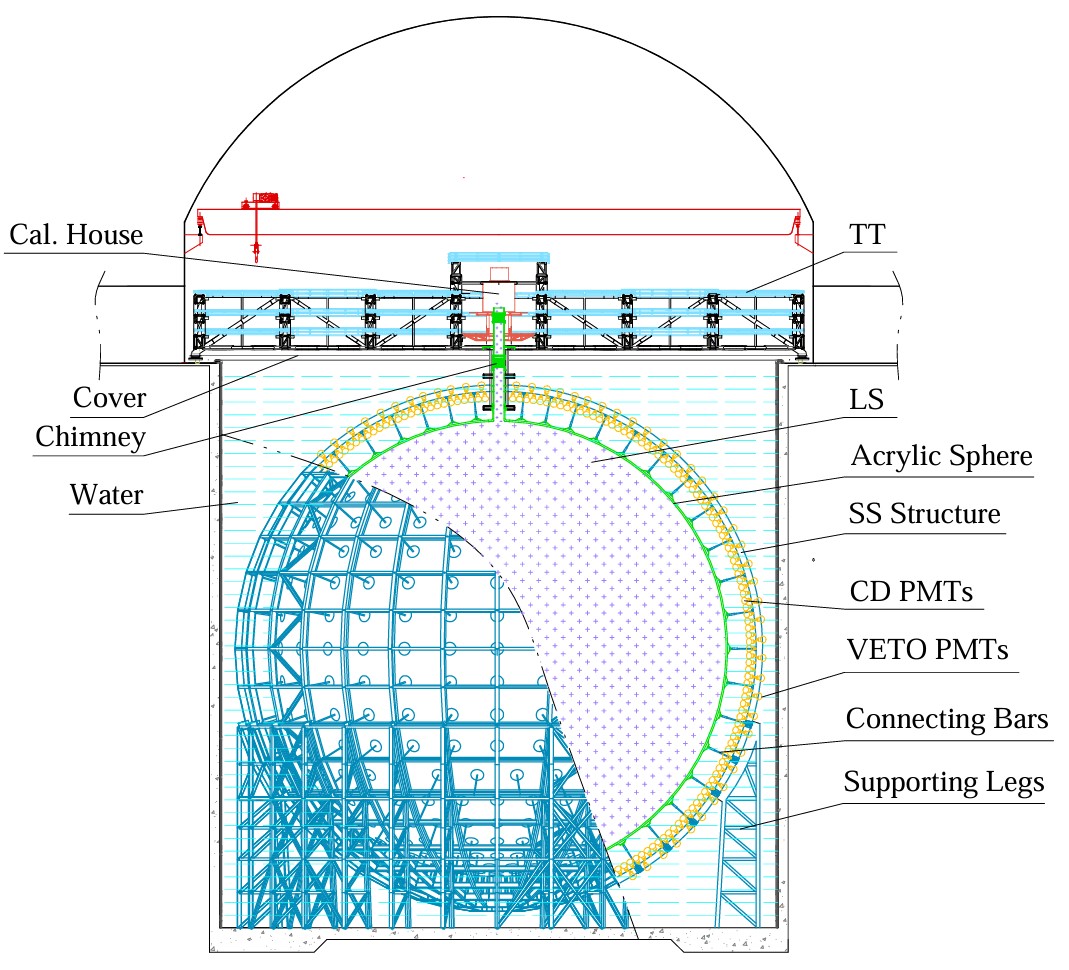}
    \caption{Schematic view of the JUNO detector\,\cite{2022PrPNP.12303927J}.}
    \label{fig:JUNO}
\end{figure}

The study of neutrinos, and of neutrino oscillations in particular, plays a special role in our understanding of particle physics and explores the frontier of our knowledge in this domain. The neutrino oscillation theory has been verified and improved through the study of solar, atmospheric, accelerator and reactor neutrinos. 
Indeed, the nonzero neutrino masses, whose only experimental evidence comes from neutrino oscillations, is our first positive indication of physics beyond the Standard Model\,\cite{GIGANTI20181}. Jiangmen Underground Neutrino Observatory (JUNO) plans to identify the neutrino mass ordering (NMO) by accurately measuring the reactor antineutrino energy spectrum. JUNO's central detector (CD) shown in Fig.\ref{fig:JUNO}, a spherical detector in a pure water pool with 20 kton liquid scintillator (LS), will use 17612 20-inch PMTs (LPMTs)\,\cite{JUNO:2022hlz} and 25600 3-inch PMTs (SPMTs)\,\cite{Cao:2021wrq, Zhang:2024okq} to achieve a photocathode coverage of more than 78\% and an excellent energy resolution of $3\%/\sqrt{E(MeV)}$. In addition, 2400 LPMTs will be installed outwards on the stainless steel grid of the CD. These PMTs and the external pool form a water Cherenkov detector, which can shield the internal environmental natural radioactivity and mark the cosmic muons together with the Top Tracker (TT)\,\cite{ABUSLEME2023168680, Luo:2024xir} as a veto detector. Besides the NMO, the large fiducial volume and the unprecedented energy resolution offer exciting opportunities for addressing many other topics in particle and astro-particle physics\,\cite{JUNO:2015zny}.

PMTs are used and studied in all typical neutrino experiments, including the Super-Kamioka\,\cite{FUKUDA2003418}, SNO\,\cite{Albanese2021TheSE}, KamLAND\,\cite{DECOWSKI201652}, Daya Bay\,\cite{DayaBay:2016ggj}, RENO\,\cite{KIM201324} and Double Chooz\,\cite{CABRERA201287}. The 20012 LPMTs of JUNO include 5000 dynode PMTs from Hamamatsu Photonics K.K. (HPK R12860)\,\cite{HPK-R12860} and 15012 microchannel plate PMTs (MCP PMTs) from Northern Night Vision Technology Ltd. (NNVT GDB6201)\,\cite{NNVT-GDB6201-note}. However, some large pulses still require further investigation with larger statistics, as the existing studies\,\cite{zhangyu-large-pulse,Zhang:2023dha} are limited to individual LPMTs.

In order to achieve the designed energy resolution of JUNO, all LPMTs have passed the acceptance test and have been potted (HV divider firmly soldered to the PMT)\,\cite{JUNOPMTsignalopt,JUNO:2022hlz}. To achieve the physics goals and experimental requirements of JUNO, the 1F3 electronics has been developed\,\cite{BELLATO2021164600,COPPI2023168255,Petitjean_2022,CERRONE2023168322}. Its key component is the Global Control Unit (GCU), with each GCU having 3 channels. Each channel is equipped with two ADCs, allowing simultaneous readout of 3 PMTs. The GCU offers a wide dynamic range of 1\,P.E.\,to 1000\,P.E.\,and exhibits excellent linear response. In total, 6681 1F3 electronics UWBoxes (Under Water box) will be used for all LPMTs in JUNO. In order to evaluate the performance of the potted LPMTs and obtain an early understanding of their response in the JUNO detector, an onboard pulse generator is used to qualify the hardware performance\,\cite{COPPI2023168255,CERRONE2023168322}. Some potted LPMTs have done the performance test using the LED light source and dark noise triggering with the 1F3 electronics UWBox in containers \#D, which are located in the JUNO Pan-Asia PMT testing and potting station in Jiangmen, Guangdong Province, China. The fundamental performance parameters of LPMTs have been investigated under single photoelectron mode triggered by LED light source\,\cite{Liu_2023}. However, the PMT self generated large pulses are a good signal to test the 1F3 electronics by its broad amplitude range and high rate as well.

This paper focuses on the combined response characteristics of LPMTs large pulses and 1F3 electronics using PMT self-generated large pulses from dark counts. A brief introduction to the container \#D testing system and 1F3 electronics is presented in Sec.\,\ref{1:system}. The large signal of LPMTs in a darkroom is shown in Sec.\,\ref{1:testing}. The characteristics and consistency of the two ranges of 1F3 electronics is discussed in Sec.\,\ref{1:comparison}. Finally, a summary is given in Sec.\,\ref{1:summary}. 

\section{Container \#D with 1F3 electronics prototype}
\label{1:system}

\begin{figure}[!htb]
    \centering
    \includegraphics[width=0.95\linewidth]{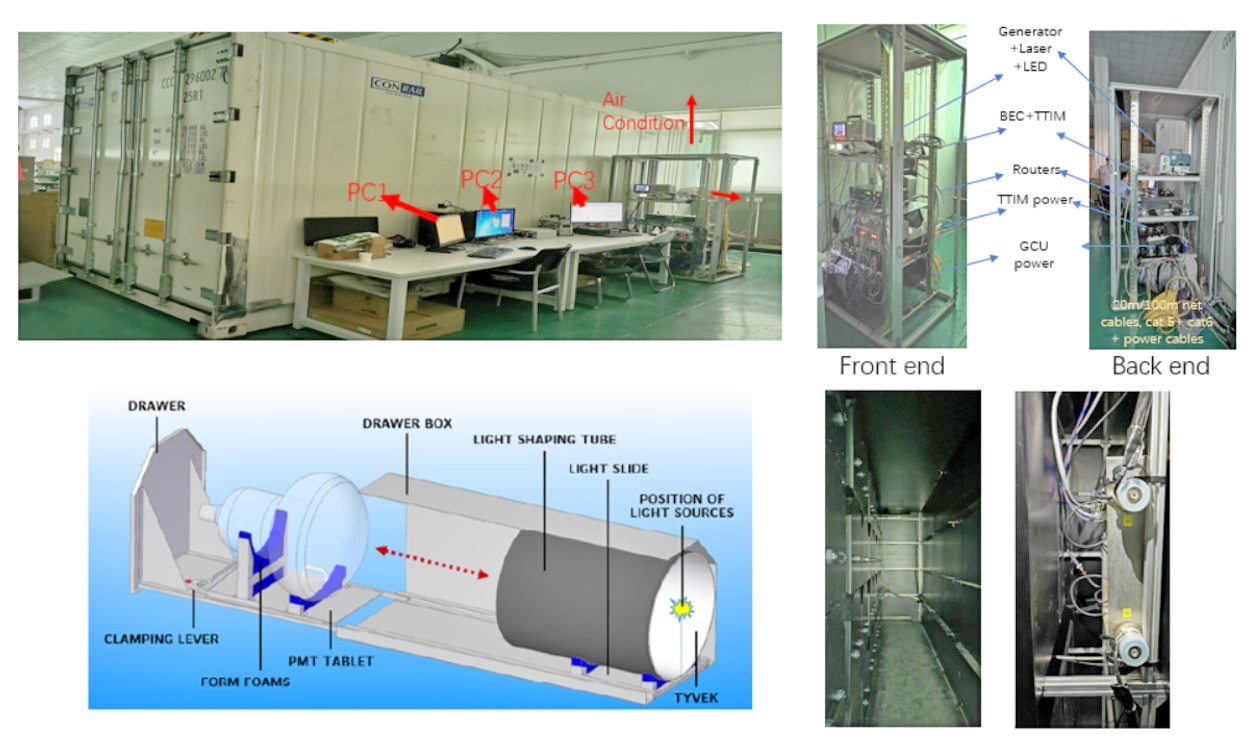}
    \caption{Top left: the outer view with control PCs of the 20-foot container \#D. Top right: the rack where the power is installed. Bottom left: the schematic view of a drawer box. Bottom right: the internal view of the container and an installed box of JUNO 1F3 electronics between the drawers.}
\label{fig:box0:C4}
\end{figure}

As shown in Fig.\ref{fig:box0:C4}\,\cite{Liu_2023}, the main body of the container \#D testing system consists of a darkroom, which is passively shielded against the EMF by a multi-layer silicon-iron shielding. It includes a total of 32 drawers, which are independent. The system is equipped with 11 UWBoxs, allowing simultaneous testing of 32 LPMTs (leaving one of the 1F3 channels unconnected). Each drawer is equipped with a LED light sources, enabling both LED-triggered and PMT self-triggered modes. Due to the potential heat generated by the electronics during long-term operation, a HVAC (Heating, Ventilation, and Air Conditioning) is installed. It is set to maintain a temperature of 23$^{\circ}$C to ensure a stable operation conditions inside the container.

\begin{figure}[!htb]
    \centering
    \includegraphics[width=0.85\linewidth]{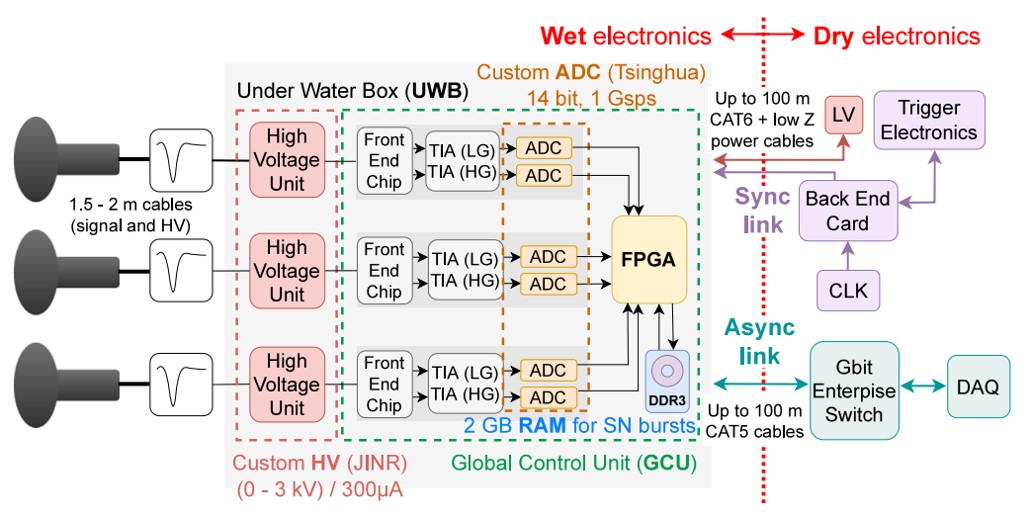}
    \caption{Schematic view of JUNO 1F3 electronics system\,\cite{COPPI2023168255}.}
\label{fig:box:1F3}
\end{figure}

As shown in Fig.\ref{fig:box:1F3}, each 1F3 electronics UWBox consists of three High Voltage Units (HVUs) and a Global Control Unit (GCU). Each HVU independently powers one LPMT and decouples the PMT signal from the high voltage. The analog signals from LPMTs are split into two parallel streams by a custom front-end chip upon reaching the GCU. These parallel streams are then fed into two pre-amplifiers, one with high gain (fine range) and the other with low gain (coarse range). Subsequently, they are converted into digital signals using individual 14 bit, 1 GSample/s custom Flash Analog-to-Digital Converters (FADC) and tagged by different ranges. After calibration, the high-gain fine range has a dynamic range of zero \,P.E. to 100 \,P.E with a resolution of 0.1 \,P.E or better. The conversion factor for this range is around 0.12\,mV/ADC\footnote{ADC as a unit represents the least significant bit (LSB).}. On the other hand, the low-gain coarse range has a dynamic range of 100 \,P.E to 1000 \,P.E with a resolution of 1 \,P.E or better. The conversion factor for this range is around 0.83\,mV/ADC. These values are based on a nominal LPMT gain of 10$^{7}$. The digital signals outputted by the FADC are further processed by a Xilinx Kintex-7 FPGA (XC7K325T), which serves as the core of the GCU. The FPGA enables functions such as local trigger generation, charge reconstruction, timestamp tagging, and temporary storage. Additionally, the GCU is equipped with a 2-GBytes DDR3 memory, which can provide a larger memory buffer in case of sudden increases in input rate. In order to gain an understanding of the response of the potted LPMTs in the detector, a subset (about 10\%, 5\% analyzed here) of potted LPMTs was tested in container \#D with the 1F3 electronics.

\section{Characterisation of the 20-inch PMT using large pulses}
\label{1:testing}

\begin{figure}[!htb]
    \centering
    \includegraphics[width=0.7\linewidth]{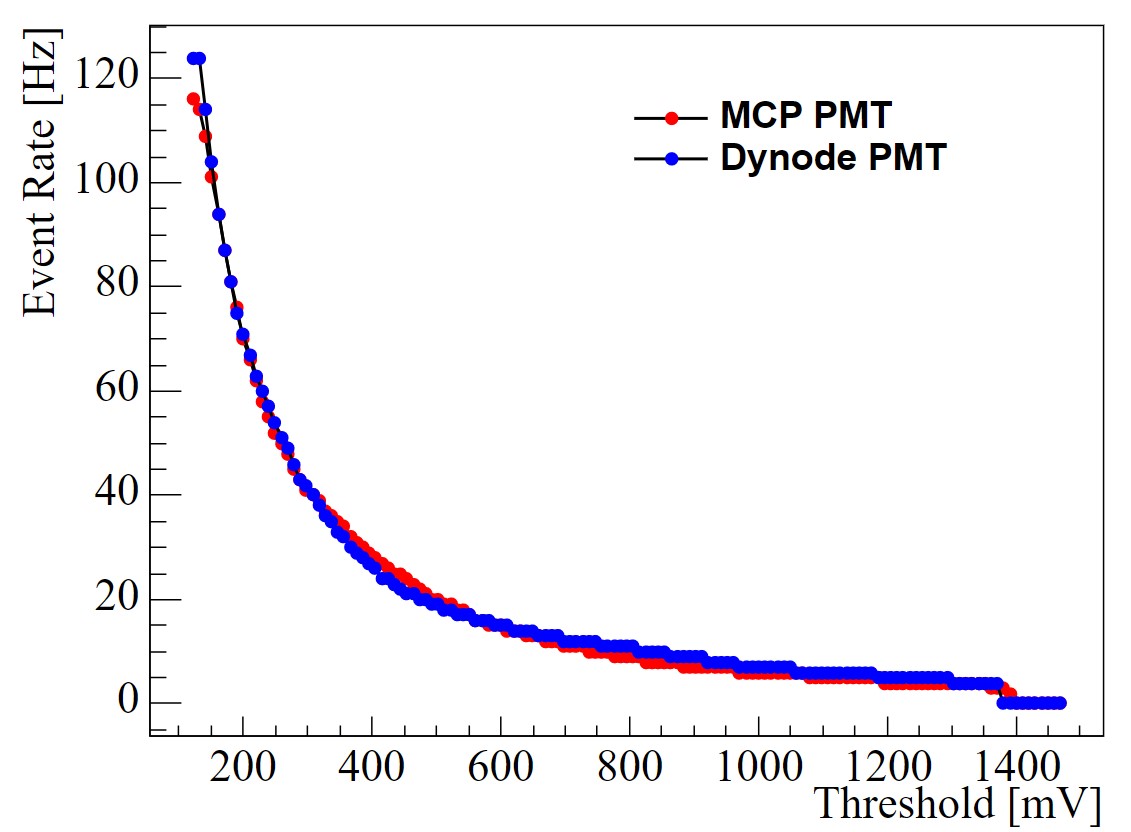}
    \caption{Threshold scanning of the event rate of one dynode PMT (blue) and one MCP PMT (red).}
\label{fig:thre_scan}
\end{figure}

The large signal testing of the PMT was conducted using a self-trigger mode, and the electronic threshold was set to 1000\,ADCs (around 120\,mV) or 4000\,ADCs (around 480\,mV) for fine range. Three electronics channels of the same GCU use the same trigger, which means when any one of them is triggered, the other two channels will synchronously acquire data too. The time window per trigger is set to ten microseconds. Each run includes 10k triggers events. Due to the high trigger threshold, only large pulses generated by the PMT itself originating from natural radioactivity and cosmic rays are recorded\,\cite{zhangyu-large-pulse,Zhang:2023dha}. A survey of large pulse rate versus threshold for fine range with a threshold set at 120mV is shown in Fig.\,\ref{fig:thre_scan}.

\subsection{Amplitude \& Charge}
\label{2:features}

\begin{figure}[!htb]
	\centering
	\includegraphics[width=0.75\linewidth]{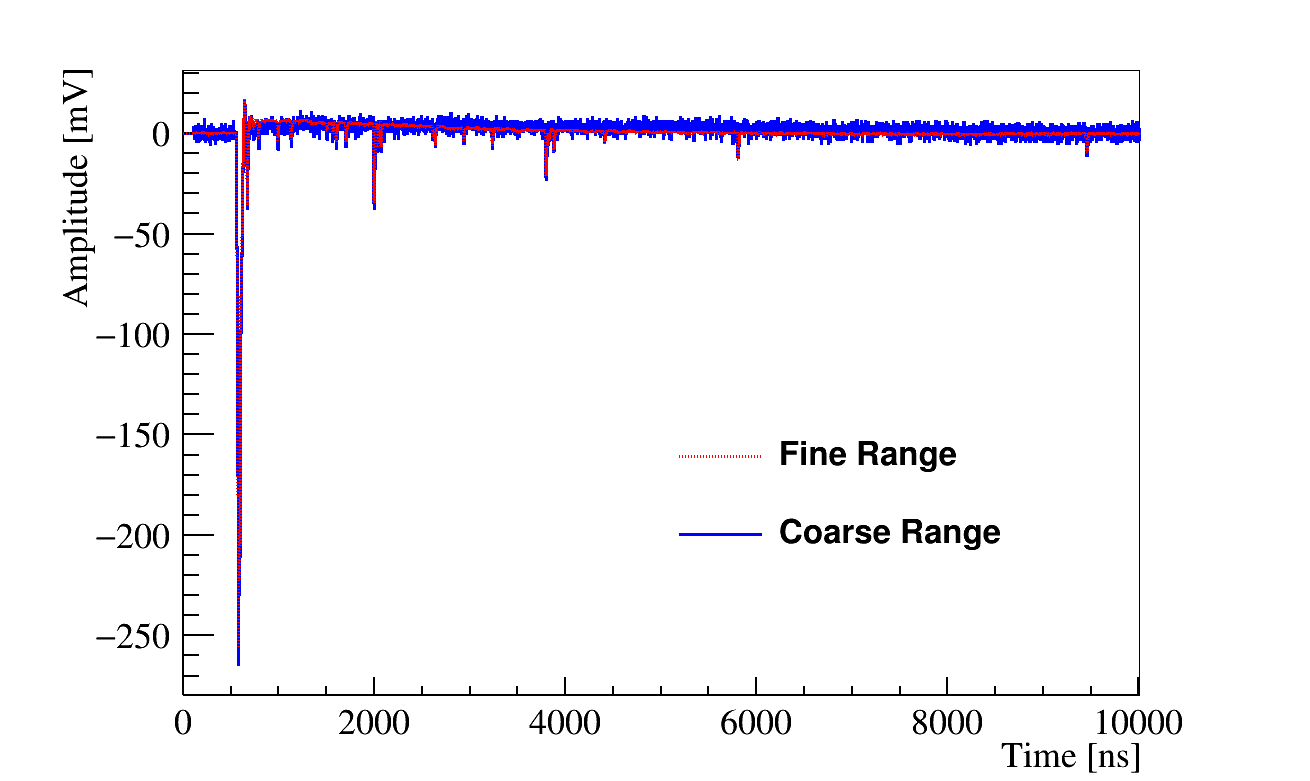}
    \caption{Exemplary baseline-corrected waveform of one recorded pulse in fine range (red histogram) and coarse range (blue histogram).}
    \label{fig:waveform}
\end{figure}

In total, 280 dynode- and 621 MCP-potted 20-inch PMTs have been tested. Fig.\,\ref{fig:waveform} shows one
recorded waveform in the fine range and coarse range for comparison. The triggered pulse located at around 600\,ns in the readout window of [0,10]\,$\mu$s.

With the test, a total of 681 PMTs were measured with a threshold of 1000\,ADCs, and the remaining ones have been tested with 4000\,ADCs. 

\begin{figure}[!htb]
    \centering
	\begin{subfigure}[c]{0.48\textwidth}
	\centering
	\includegraphics[width=\linewidth]{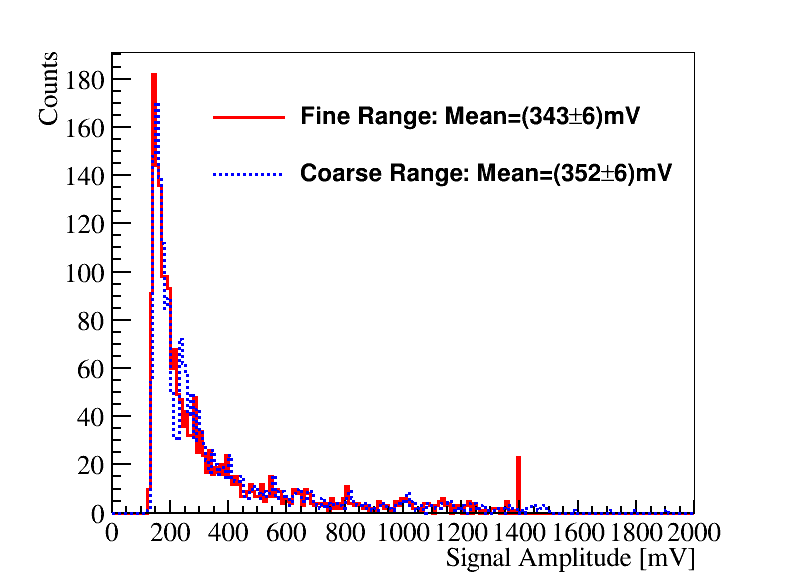}
	\caption{Amplitude}
	\label{fig:features:amplitude01}
	\end{subfigure}
	\begin{subfigure}[c]{0.48\textwidth}
	\centering
	\includegraphics[width=\linewidth]{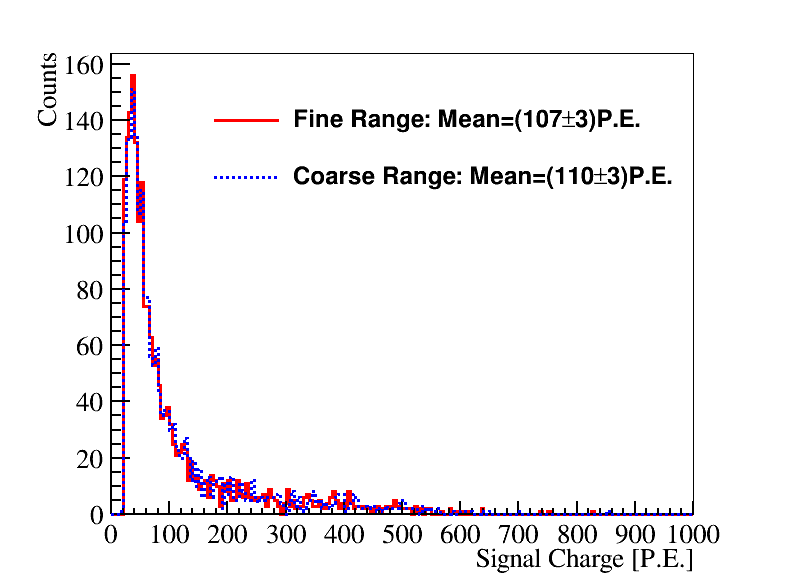}
	\caption{Charge}
	\label{fig:features:charge01}
	\end{subfigure}
    \caption{The amplitude (a) and charge (b) distributions of a NNVT MCP PMT for the fine range (red histogram) 
and the coarse range (blue histogram).}
    \label{fig:features}
\end{figure}

There are many methods for waveform reconstruction\,\cite{HUANG201848, Tang_2025}. Here, the direct integration approach is adopted for charge extraction due to its fast processing speed and low resource consumption. Fig.\,\ref{fig:features:amplitude01} present the distributions of amplitude and charge for a MCP PMT. Given that the set threshold is 1000\,ADCs, the amplitude starts from 120\,mV. It can be seen that the upper limit of the measurement of the fine range is around 1400\,mV. If it exceeds the dynamic range of the fine range, it is necessary to use the coarse range for testing. 
The rate of large pulses per PMT was obtained from each individual GCU channel of fine range by applying electronic threshold on amplitude. From Fig.\,\ref{fig:features:charge01}, it can be observed that the charge distribution of large pulse has a wide distribution too, with the a mean value of approximately 100\,P.E. Up to the value of 400\,P.E., the charge distributions of the two ranges are largely consistent with each other. 

\subsection{Timing}
\label{2:timng}

\begin{figure}[!htb]
    \centering
	\begin{subfigure}[c]{0.43\textwidth}
	\centering
	\includegraphics[width=\linewidth]{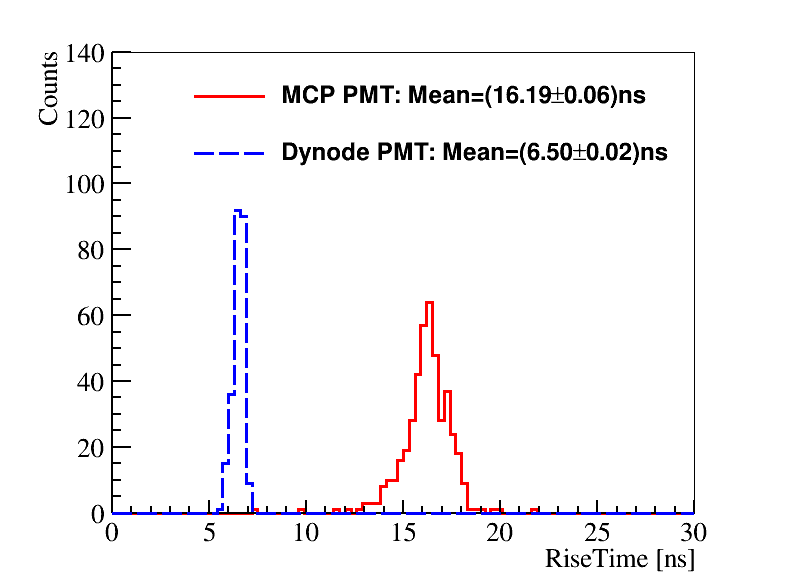}
	\caption{Rise-time}
	\label{fig:features:rise}
	\end{subfigure}
	\begin{subfigure}[c]{0.43\textwidth}
	\centering
	\includegraphics[width=\linewidth]{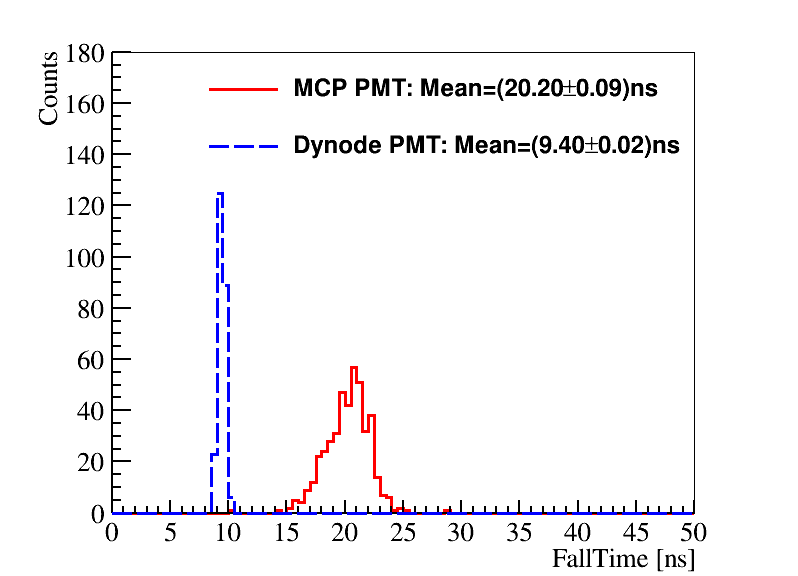}
	\caption{Fall-time}
	\label{fig:features:fall}
	\end{subfigure}
   \begin{subfigure}[c]{0.43\textwidth}
	\centering
	\includegraphics[width=\linewidth]{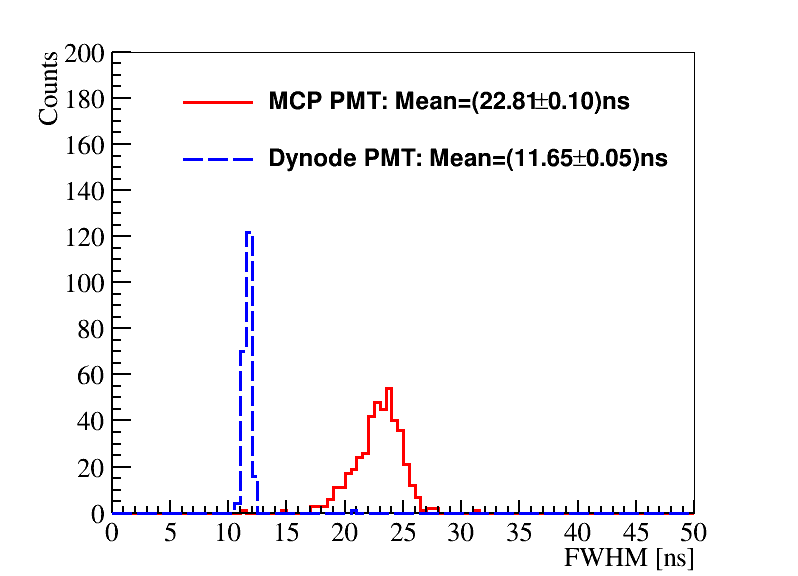}
	\caption{FWHM}
	\label{fig:features:fwhm}
	\end{subfigure}
	\begin{subfigure}[c]{0.43\textwidth}
	\centering
	\includegraphics[width=\linewidth]{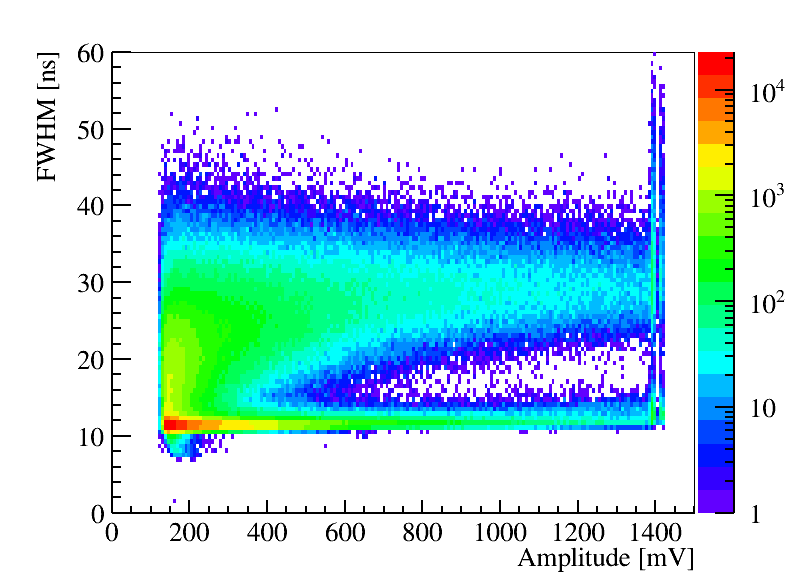}
	\caption{Amplitude vs. FWHM}
	\label{fig:features:time:2D}
	\end{subfigure}
    \caption{The rise-time (a), falltime (b) and FWHM (c) distributions of MCP PMT (red histogram) and dynode PMT (blue histogram) with fine range. Besides, amplitude vs. FWHM (d) is given.}
    \label{fig:features:time}
\end{figure}

The rise-time, fall-time and FWHM (Full width of half maximum) of the large pulses are shown in Fig.\,\ref{fig:features:time} with 1000\,ADCs threshold of DAQ. The rise-time is the duration for the waveform to rise from 10\% to 90\% of its maximum amplitude, while the fall-time is the duration for it to fall from 90\% to 10\%. The typical rise-time is 16.2\,ns for MCP PMTs, and 6.5\,ns for dynode PMTs. The typical value of fall-time is 20.2\,ns for MCP PMTs, and 9.4\,ns for dynode PMTs. The typical value of FWHM is 22.8\,ns for MCP PMTs, and 11.6\,ns for dynode PMTs. A comparison with the data from Reference\,\cite{Liu_2023}, which used the LED light source for single photoelectron level measurements, reveals that the three timing parameters of the dynode PMT show minimal variation, whereas those of the MCP PMT demonstrate pronounced differences. Fig.\,\ref{fig:features:time:2D} shows the correlation between signal amplitude and FWHM, with the upper and lower sections corresponding to the MCP PMT and dynode PMT, respectively. The distribution indicates that the FWHM of the dynode PMT stays consistently narrow and stable with increasing signal amplitude, whereas the FWHM of the MCP PMT initially has a wider distribution but gradually narrows as the signal amplitude increases.

\subsection{Rate}
\label{2:event_rate}

\begin{figure}[!htb]
    \centering
    \includegraphics[width=0.75\linewidth]{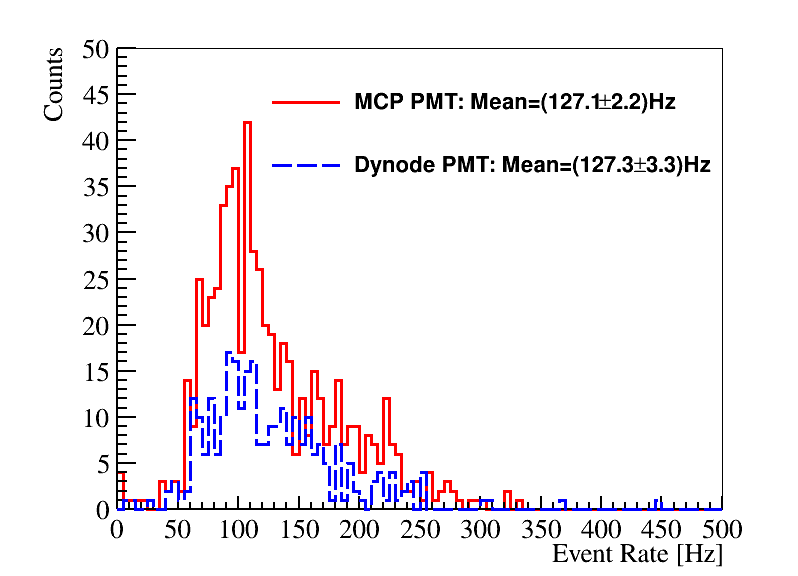}
    \caption{The event rate of large pulse for MCP PMTs (red histogram) and dynode PMTs (blue histogram).}
    \label{fig:rate}
\end{figure}

The rate of PMT self generated large pulses is defined as the event number divided by the test time with the threshold 1000\,ADCs. For each PMT, the test time is the time period between the first trigger and the last trigger, and the number of events is the number of signals whose samples exceed the set threshold during the test time. According to the calculation, the distribution of the large signal event rate detected by the two types of PMT is essentially equivalent. As shown in Fig.\,\ref{fig:rate}, the average rate of large signal is 127\,Hz only with the 1000\,ADCs threshold. In this reference\,\cite{zhangyu-large-pulse}, the event rates measured with individual LPMTs are around 100\,Hz (included in Fig.\,\ref{fig:rate}), and it is explained that the main sources of events are cosmic rays, natural radioactivity, and so on.

\subsection{Correlated pulses}
\label{2:afterpulse}

\begin{figure}[!htb]
    \centering
    \begin{subfigure}[c]{0.48\textwidth}
	\centering
	\includegraphics[width=\linewidth]{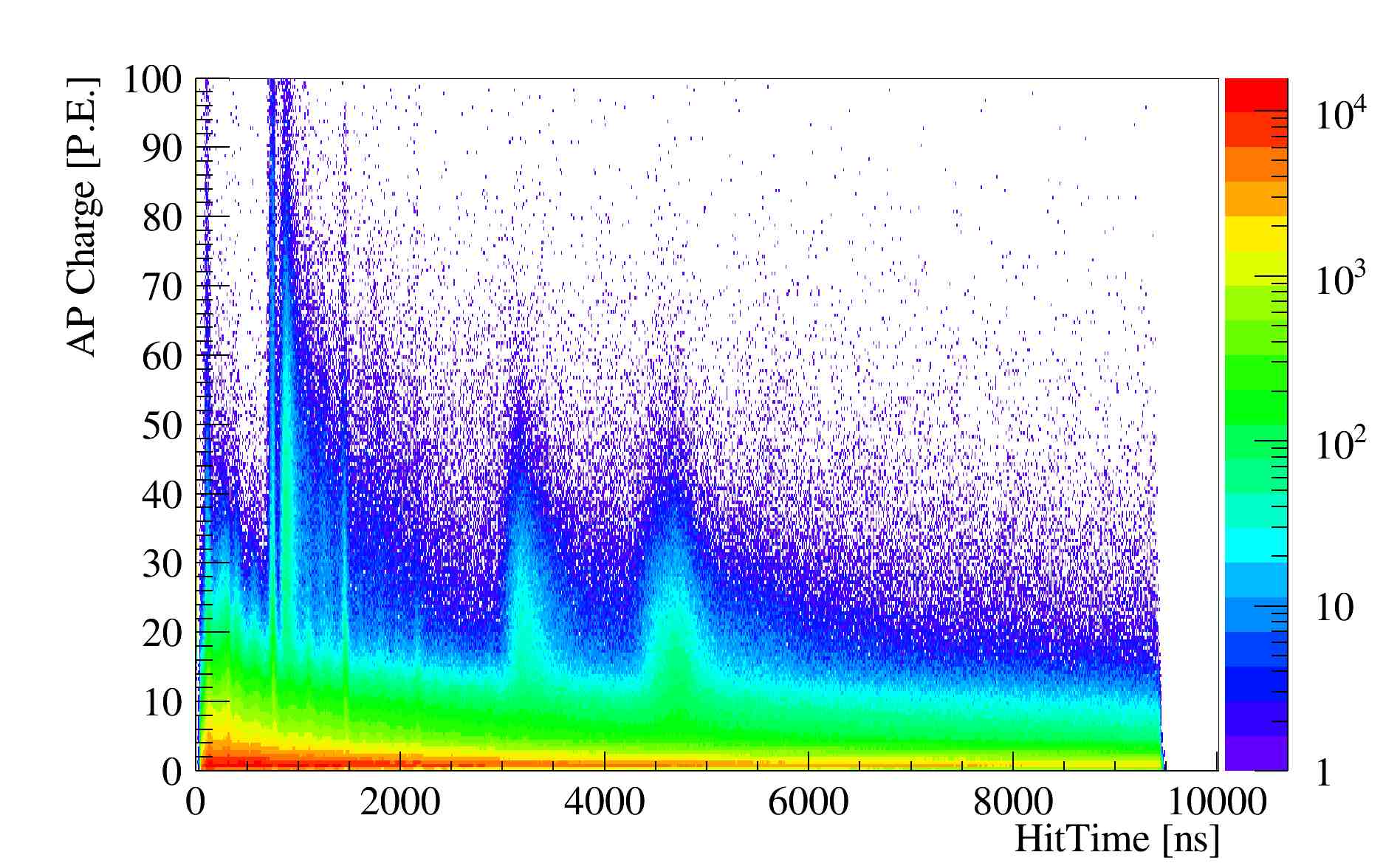}
	\caption{MCP PMT}
	\label{fig:APT_Q:mcp}
	\end{subfigure}
	\begin{subfigure}[c]{0.48\textwidth}
	\centering
	\includegraphics[width=\linewidth]{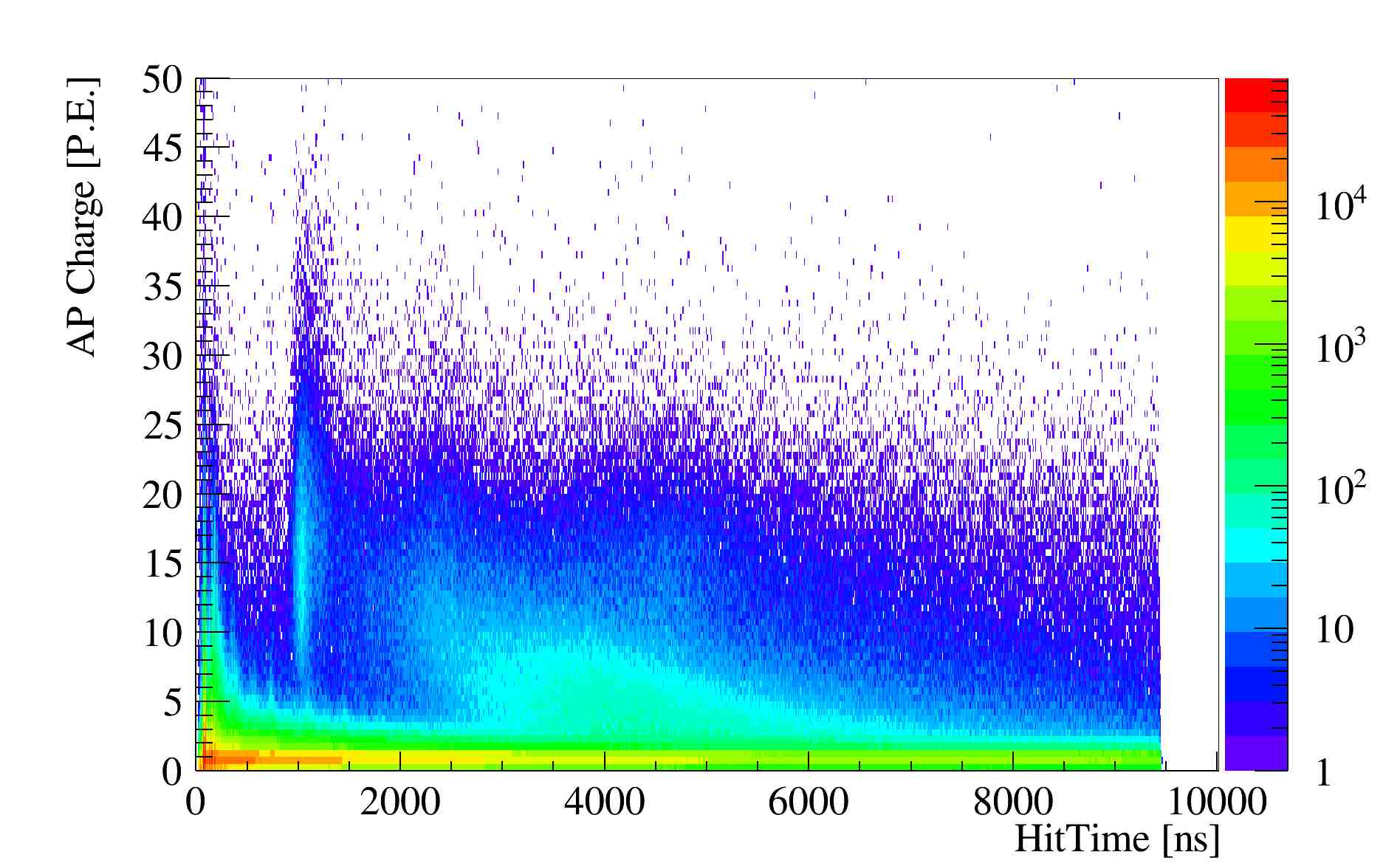}
	\caption{Dynode PMT}
	\label{fig:APT_Q:ham}
	\end{subfigure}
    \caption{Time and charge of the correlated pulses (AP) after the PMT self-generated large pulses of MCP PMT (a) and dynode PMT (b).}
    \label{fig:APT_Q}
\end{figure}

Due to the spread distribution on amplitude of the PMT self-generated large pulses, we can use them to investigate the correlated pulses of PMTs. In the recorded data, it can be observed that
signal pulses (PMT self-generated large pulses) are often followed by after pulses (APs), which can be divided into fast APs on the nanosecond scale and slow APs on the microsecond scale. Fast APs are primarily triggered by photoelectrons back-scattered from the first dynode. These photoelectrons decelerate towards the photocathode under the electric field, then accelerate back towards the dynode, thus generating APs close to the signal pulse. Slow APs can be triggered by ionized residual gas. During the acceleration of photoelectrons from the photocathode to the dynode, residual gas molecules may be ionized. The positively charged ions accelerate towards the photocathode, generating secondary electrons, which finally trigger slow APs farther from the signal pulse. Fig.\,\ref{fig:APT_Q} illustrates the 2-D distribution of the after-pulse arrival time (time interval from the signal pulse) versus their charges, where the primary pulse (PMT self-generated large pulse) is located at around 545\,ns in the window, and the charge in photoelectron (P.E.) is calculated according to the peak charge\,\cite{pmtgain_Zhang_2021}. It shows a typical AP feature of 20-inch PMT on timing as shown in \cite{WU2021165351-MCP-afterpulse,rong-afterpulse}, where all the typical peaks in charge are located on top of a plateau of few P.E.\,for both MCP PMT and dynode PMT. The first nanosecond-scale region should be the fast AP component, and the following several clustered regions should be the slow AP component caused by different gas ions, with the rest being due to other reasons and background. This study targets all slow AP components after 500 ns.

\begin{figure}[!htb]
    \centering
    	\begin{subfigure}[c]{0.48\textwidth}
	\centering
	\includegraphics[width=\linewidth]{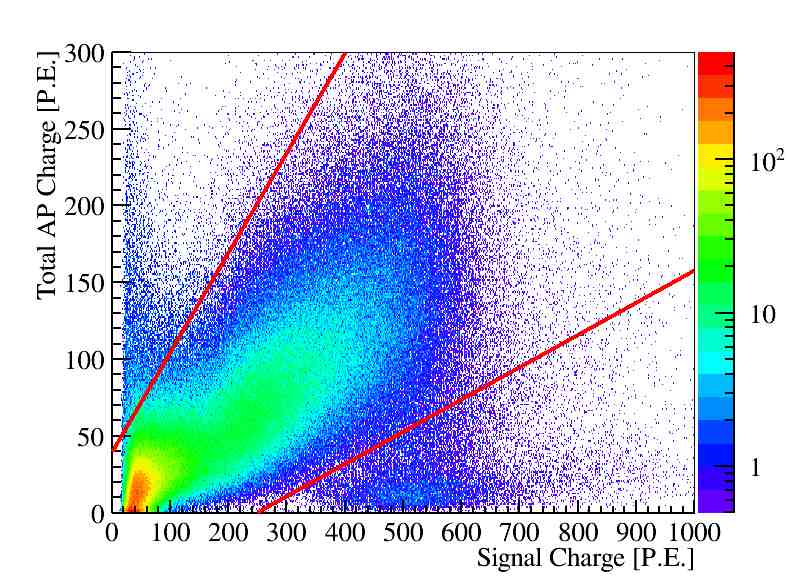}
	\caption{Charge correlation between primary and after pulses for MCP PMT}
	\label{fig:SignalQ_APTQ:mcp}
	\end{subfigure}
       \begin{subfigure}[c]{0.48\textwidth}
	\centering
	\includegraphics[width=\linewidth]{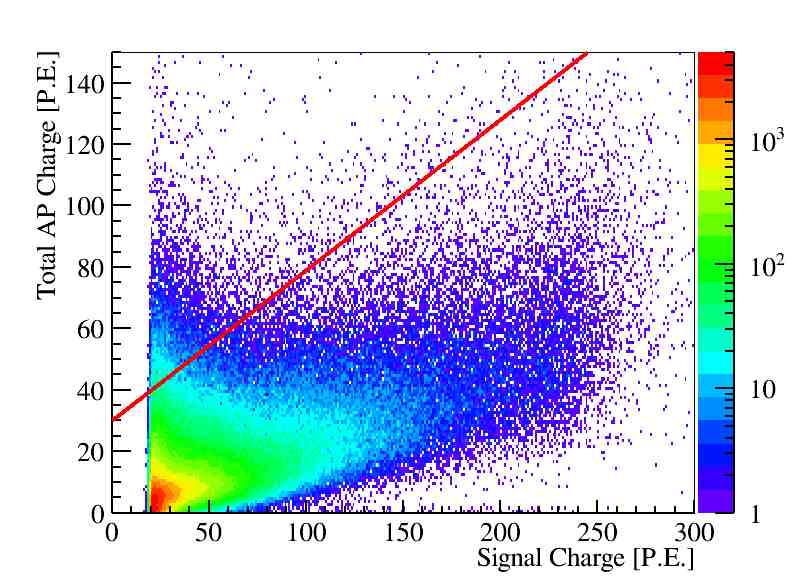}
	\caption{Charge correlation between primary and after pulses for Dynode PMT}
	\label{fig:SignalQ_APTQ:ham}
	\end{subfigure}
    	\begin{subfigure}[c]{0.48\textwidth}
	\centering
	\includegraphics[width=\linewidth]{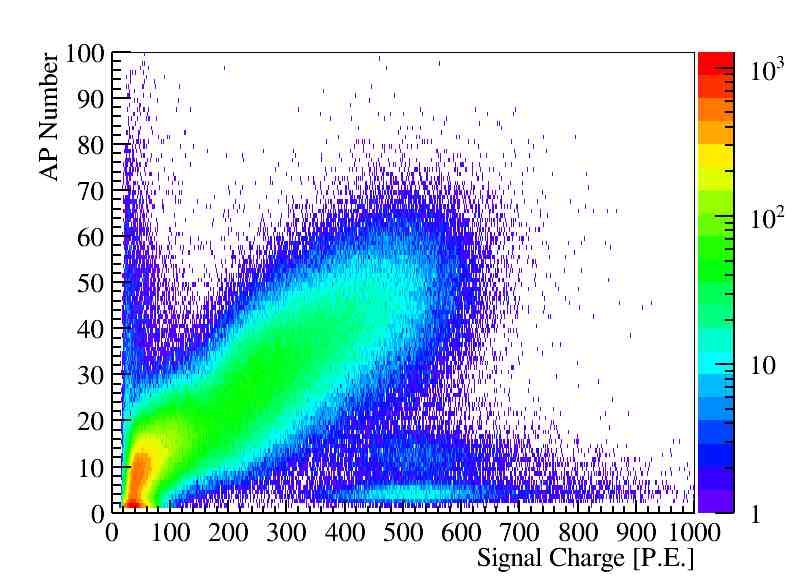}
	\caption{AP count vs.~primary charge for MCP PMT}
	\label{fig:SignalQ_APN:mcp}
	\end{subfigure}
	\begin{subfigure}[c]{0.48\textwidth}
	\centering
	\includegraphics[width=\linewidth]{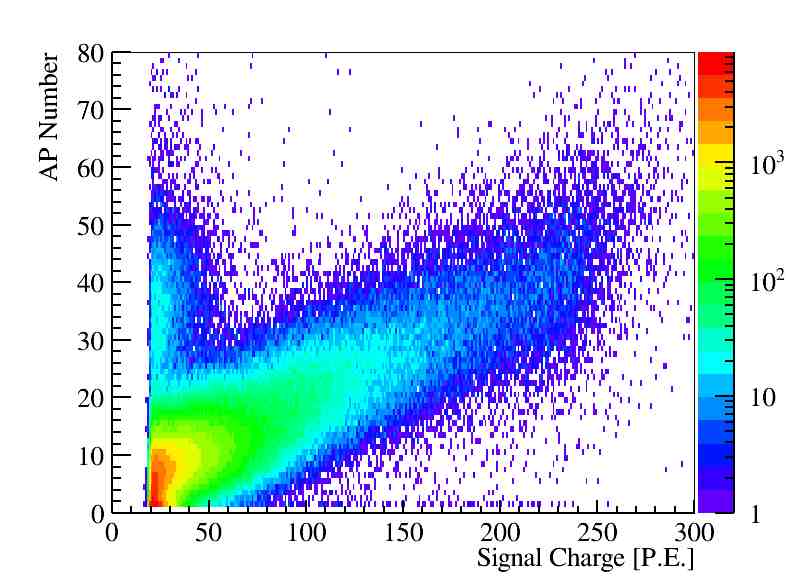}
	\caption{AP count vs.~primary charge for Dynode PMT}
	\label{fig:SignalQ_APN:ham}
	\end{subfigure}
    \caption{2-D distributions of AP count and charge versus the primary pulses for MCP PMT and dynode PMT.}
    \label{fig:SignalQ_APTQ_APN}
\end{figure}

Fig.\,\ref{fig:SignalQ_APTQ:mcp} and \ref{fig:SignalQ_APTQ:ham} further investigate the behavior of the total APs charge versus the charge of the primary pulses. It is easy to identify three regions (top left, middle, bottom right) on the 2-D distribution of MCP PMT (Fig.\,\ref{fig:SignalQ_APTQ:mcp}) as divided by the two red lines. The region of top left shows a larger charge of AP (related to the AP around 500\,ns after the primary pulses) with a smaller primary pulse charge (most of them less than 50\,P.E.). The region of bottom right is more like a standalone peak from the primary pulses with around 500\,P.E. The middle region is characterized by a proportionality between the total charge of AP and primary pulses as described in previous studies\,\cite{WU2021165351-MCP-afterpulse,rong-afterpulse}. Conversely, the dynode PMT (Fig.\,\ref{fig:SignalQ_APTQ:ham}) shows only two regions as the bottom right one disappears. Fig.\,\ref{fig:SignalQ_APN:mcp} and \ref{fig:SignalQ_APN:ham} show the total APs count per primary pulse versus the charge of the primary pulses. They are showing a similar distribution and features as the charge distributions three regions for MCP PMT and only two regions for dynode PMT. 

\begin{figure}[!htb]
    \centering
    \begin{subfigure}[c]{0.48\textwidth}
	\centering
	\includegraphics[width=\linewidth]{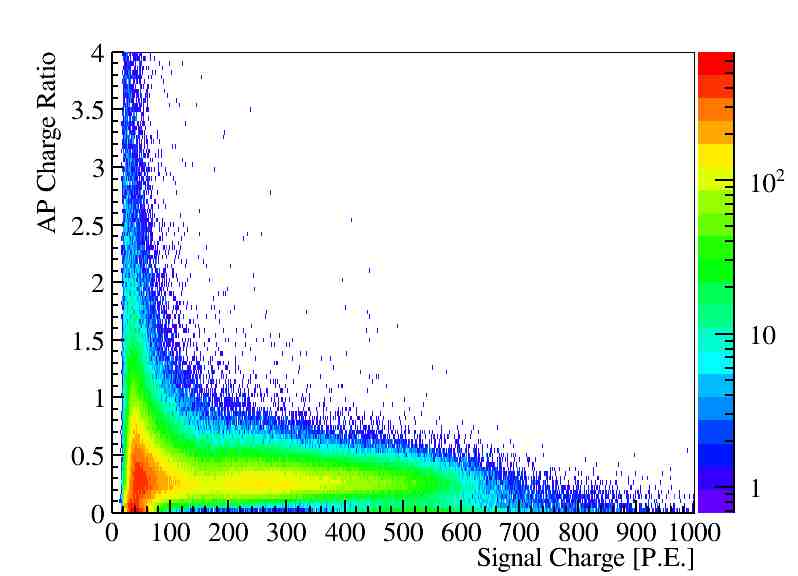}
	\caption{MCP PMT}
	\label{fig:SignalQ_APQRatio:mcp}
	\end{subfigure}
	\begin{subfigure}[c]{0.48\textwidth}
	\centering
	\includegraphics[width=\linewidth]{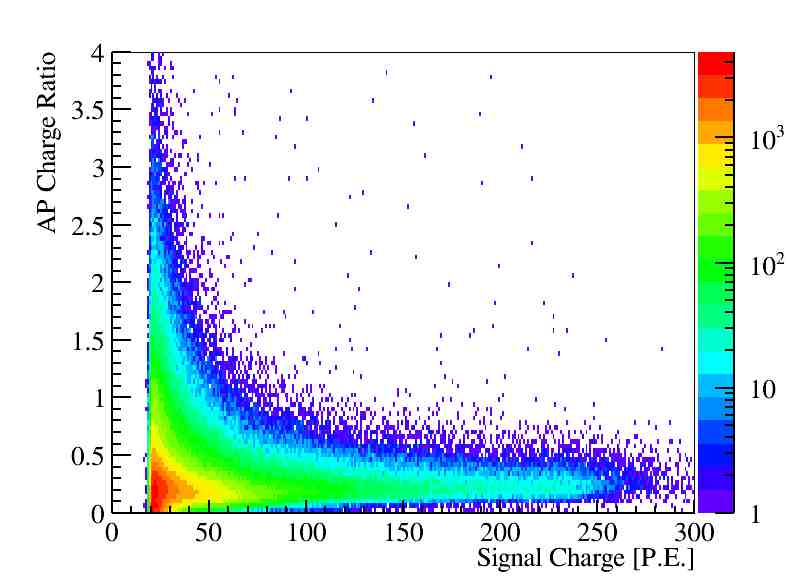}
	\caption{Dynode PMT}
	\label{fig:SignalQ_APQRatio:ham}
	\end{subfigure}
    \caption{2-D distributions of the charge ratio of AP to primary pulse for MCP PMT (a) and dynode PMT (b).}
    \label{fig:SignalQ_APQRatio}
\end{figure}

Concerning the charge ratio of the AP to its primary pulse, Fig.\,\ref{fig:SignalQ_APQRatio} shows the calculation results of all the primary pulses on Fig.\,\ref{fig:SignalQ_APTQ_APN}. It can be  deducted that: the AP charge ratio is much higher when the charge of primary pulse is small comparing to larger primary pulses, which should be related to the top left region event on Fig.\,\ref{fig:SignalQ_APTQ_APN}. The AP charge ratio for larger primary pulses is also larger than the previous measurement in literature\,\cite{WU2021165351-MCP-afterpulse,rong-afterpulse}, which is contributed by the strong plateau on Fig.\,\ref{fig:APT_Q}.

\begin{figure}[!htb]
    \centering
    \includegraphics[width=0.75\linewidth]{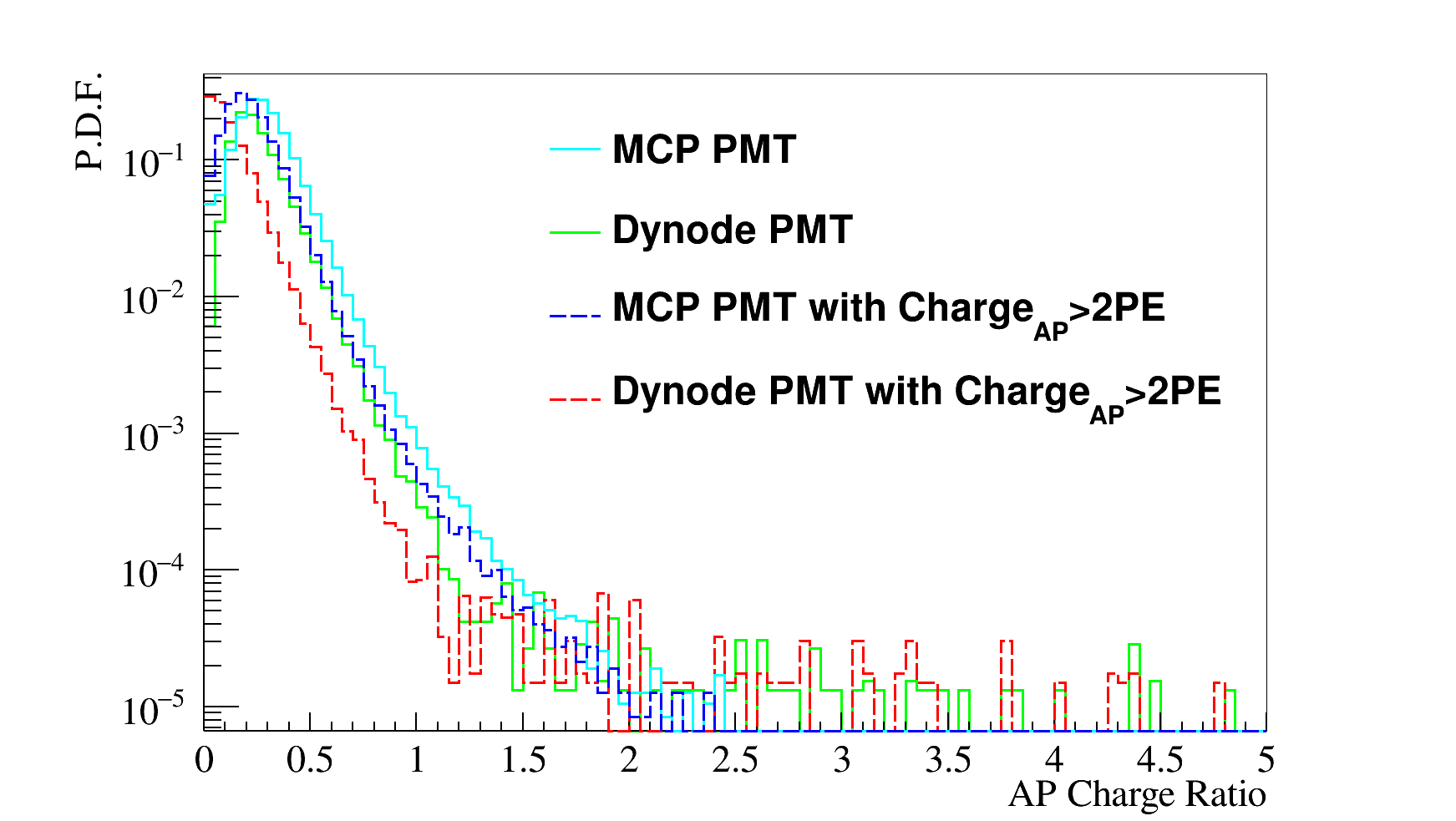}
    \caption{AP charge ratio to primary pulse.}
    \label{fig:afterpulse}
\end{figure}

Fig.\,\ref{fig:afterpulse} shows the 1-D distribution of the AP charge ratio for two types of PMT when the primary pulse is larger than 100\,P.E. with or without considering the few photoelectron plateau in time. With the threshold of 2\,P.E., the average AP charge ratio of the dynode PMT decreases from 25.9\% to 12.9\%, and that of the MCP PMT from 28.9\% to 22.6\%. Compared with the results with the scanning station system in this article\,\cite{rong-afterpulse} , the dynode PMT result is consistent, but the MCP PMT result is still much larger as in our tests we considered all slow AP contributions rather than only the gas ion one.

\section{Features of the 1F3 electronics using 20-inch PMT large pulses}
\label{1:comparison}

\subsection{Noise}
\label{2:noise}
\begin{figure}[!htb]
	\centering
	\includegraphics[width=0.6\linewidth]{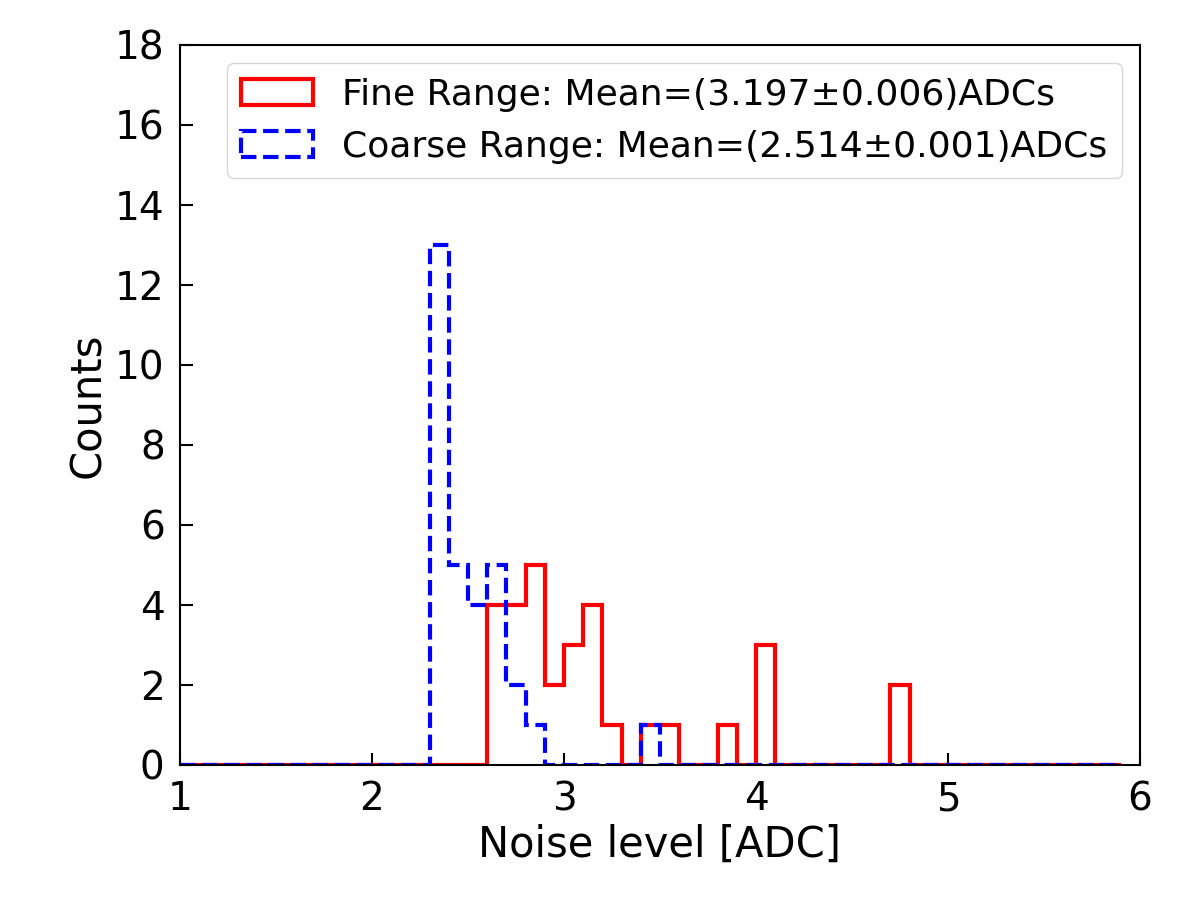}
    \caption{Noise of 1F3 electronics prototype with fine range (red histogram) and coarse range (blue histogram).}
    \label{fig:noise}
\end{figure}

As shown in Fig.\ref{fig:noise}, the average noise level of the 32 electronic channels in Container \#D is evaluated by the standard deviation of the first 100 ns of the signal window. If there is a pulse in this section, the pulse is removed before calculating the mean value. With this procedure we can estimate a noise level of around 3.2\,ADCs on average for fine range (around 0.4\,mV, around 0.05\,P.E.@1e7 gain), and around 2.5\,ADCs on average for coarse range (around 2.1\,mV, around 0.5\,P.E.@1e7 gain).

\subsection{Dynamic range on Amplitude vs Charge}
\label{2:charge}

\begin{figure}[!htb]
    \centering
    \begin{subfigure}[c]{0.48\textwidth}
	\centering
	\includegraphics[width=\linewidth]{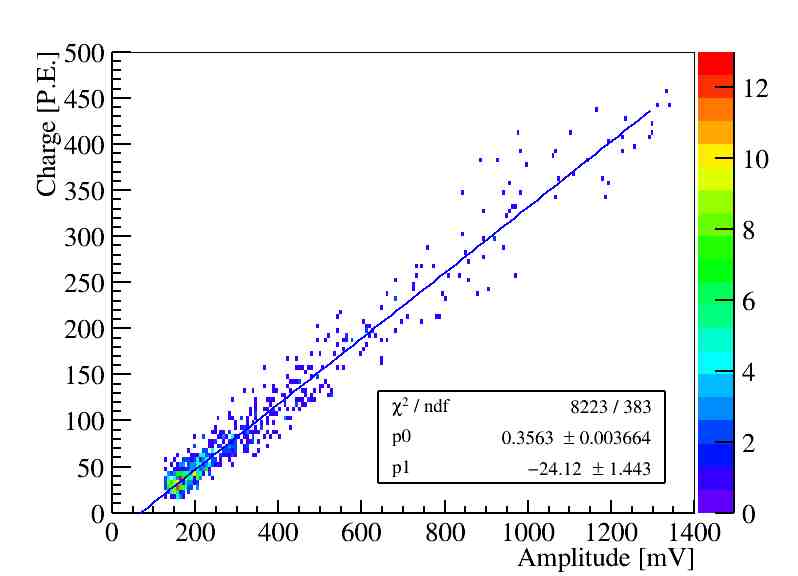}
	\caption{Fine range}
	\label{fig:charge:range1_amp_Q}
	\end{subfigure}
	\begin{subfigure}[c]{0.48\textwidth}
	\centering
	\includegraphics[width=\linewidth]{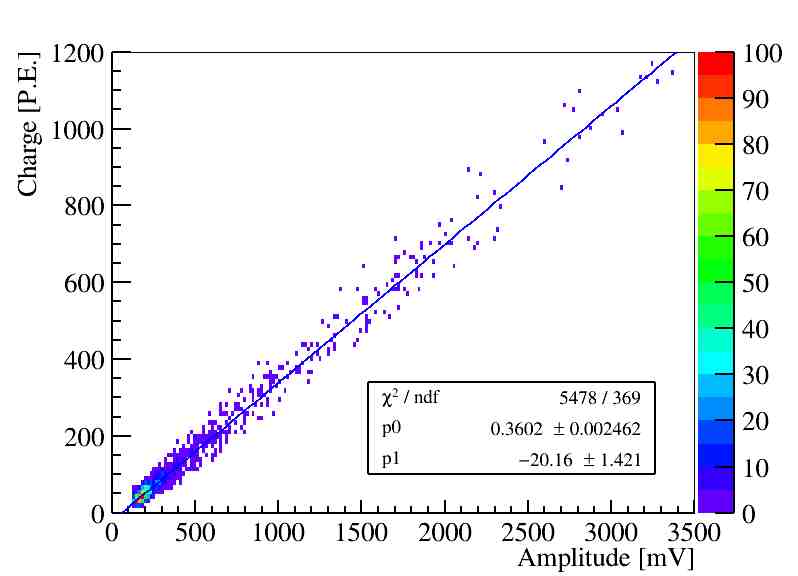}
	\caption{Coarse range}
	\label{fig:charge:range0_amp_Q}
	\end{subfigure}
    \caption{The relationship between charge and amplitude of fine range (a) and coarse range (b).}
    \label{fig:charge}
\end{figure}

\begin{figure}[!htb]
    \centering
    \includegraphics[width=0.75\linewidth]{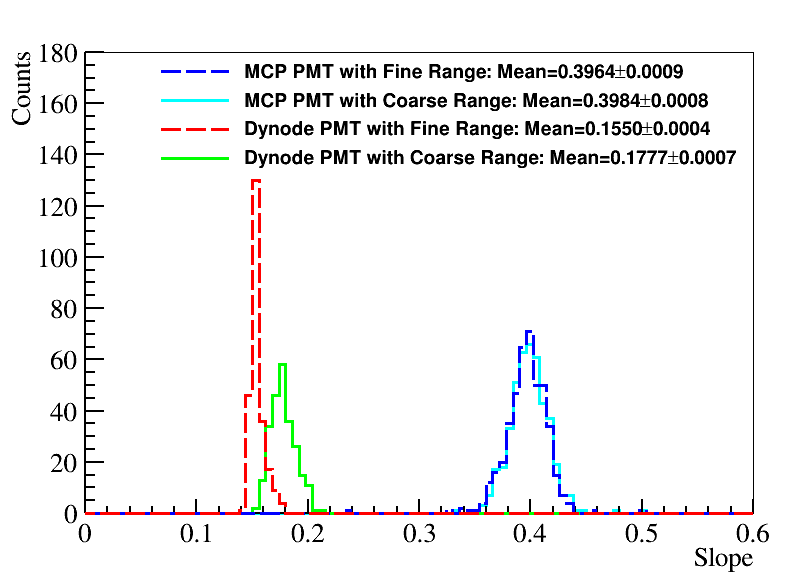}
    \caption{The slope of amplitude-charge fitting for MCP PMT and dynode PMT with two ranges.}
    \label{fig:amp_charge_slope}
\end{figure}

It is well known that the charge of a PMT pulse is proportional to its amplitude. Therefore, a linear function is used to model the relationship between charge and amplitude of the PMT pulse for both ranges. One example for a MCP PMT is shown in Fig.\,\ref{fig:charge:range1_amp_Q} and Fig.\,\ref{fig:charge:range0_amp_Q}. The fitted slopes for all tested PMTs are shown in Fig.\,\ref{fig:amp_charge_slope}.
The typical slope for dynode PMTs (HPK) for both fine range and coarse ranges is around 0.16, while it is around 0.40 for MCP PMTs. The difference between the two kinds of PMTs is mainly from their pulse shape as previously discussed.

\begin{figure}[!htb]
    \centering
        \begin{subfigure}[c]{0.43\textwidth}
	\centering
	\includegraphics[width=\linewidth]{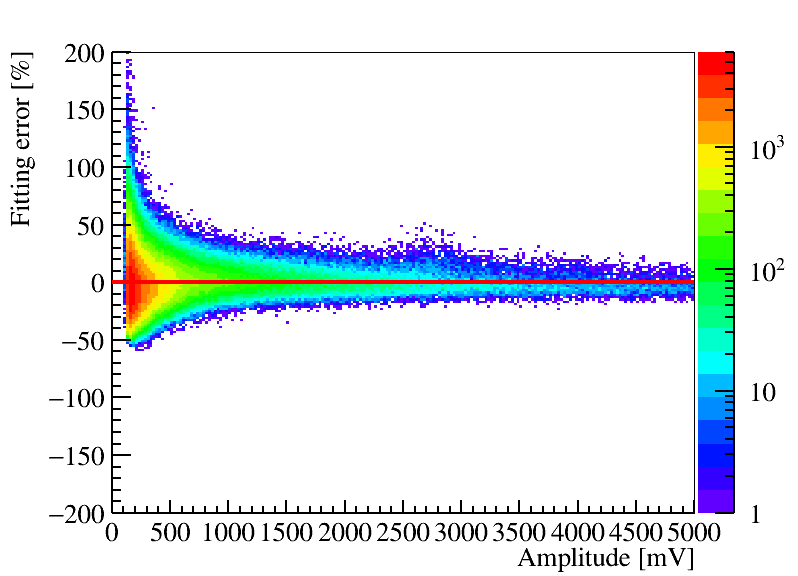}
	\caption{MCP PMT}
	\label{fig:charge:range0_AQfiterror_mcp}
	\end{subfigure}
	\begin{subfigure}[c]{0.43\textwidth}
	\centering
	\includegraphics[width=\linewidth]{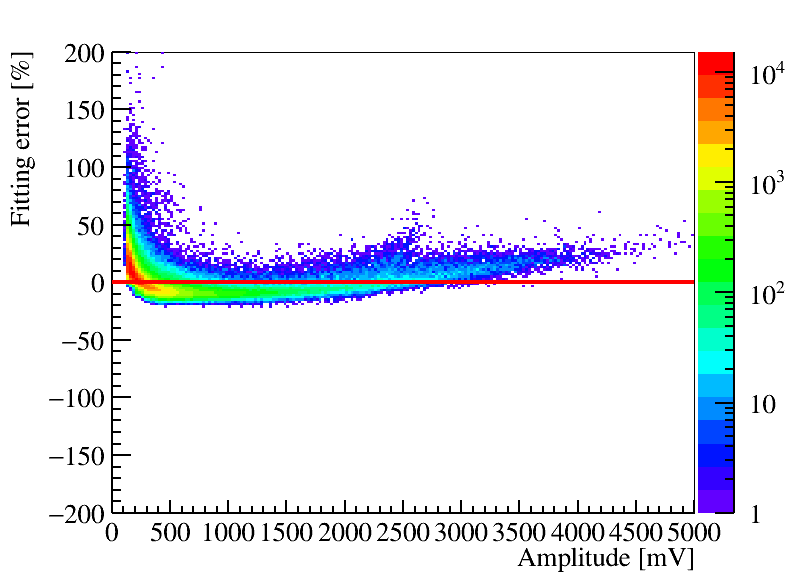}
	\caption{Dynode PMT}
	\label{fig:charge:range0_AQfiterror_ham}
	\end{subfigure}
    \caption{The change in the relative fitting error of charge and amplitude with amplitude for MCP PMT (a) and dynode PMT (b).}
    \label{fig:charge:err}
\end{figure}

Furthermore, the relative error of the fit is calculated as the ratio of the difference between the measured values and the fitted values to the fitted values. The relative fitting error for the two types of PMTs for coarse range are presented in Fig.\,\ref{fig:charge:range0_AQfiterror_mcp} and Fig.\,\ref{fig:charge:range0_AQfiterror_ham}. 
It means that the charge variation for the same amplitude of the MCP PMT is approximately 40\%, which is much larger than that of the dynode PMT. This factor also can reflect the resolution of charge measurement, and this difference may be attributed to the special long-tail structure of the MCP PMT charge distribution\,\cite{pmtgain_Zhang_2021}. It also can be seen that within the measurement range of the fine range, both types of PMTs exhibit a good linear relationship between amplitude and charge. However, as the amplitude continues to increase, the linear relationship of the MCP PMT remains intact, while the dynode PMT starts to exhibit a non-linear relationship between amplitude and charge.

\subsection{Pulse Overshoot with 1F3 electronics}
\label{2:overshoot}

\begin{figure}[!htb]
    \centering
        \begin{subfigure}[c]{0.43\textwidth}
	\centering
	\includegraphics[width=\linewidth]{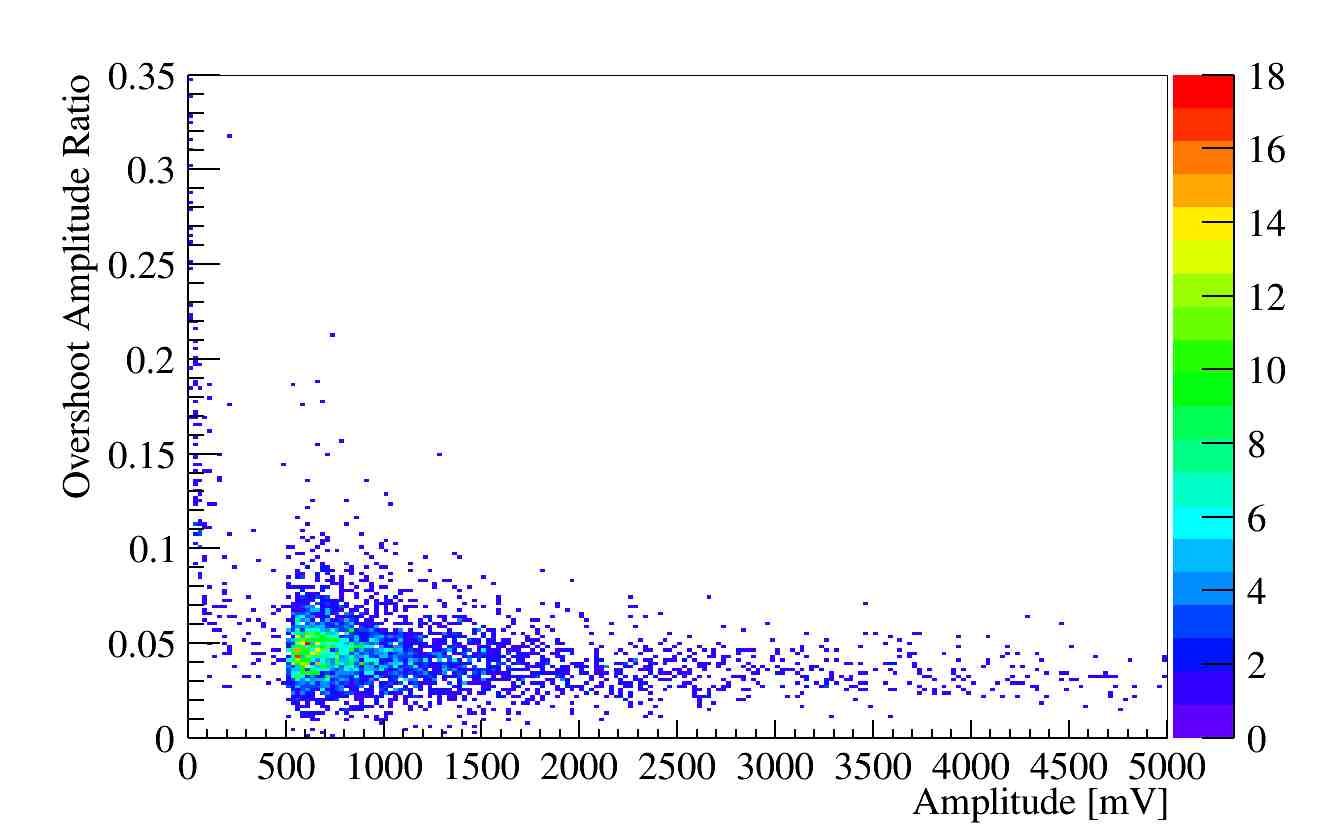}
	\caption{MCP PMT}
	\label{fig:ampratio_mcp}
	\end{subfigure}
	\begin{subfigure}[c]{0.43\textwidth}
	\centering
	\includegraphics[width=\linewidth]{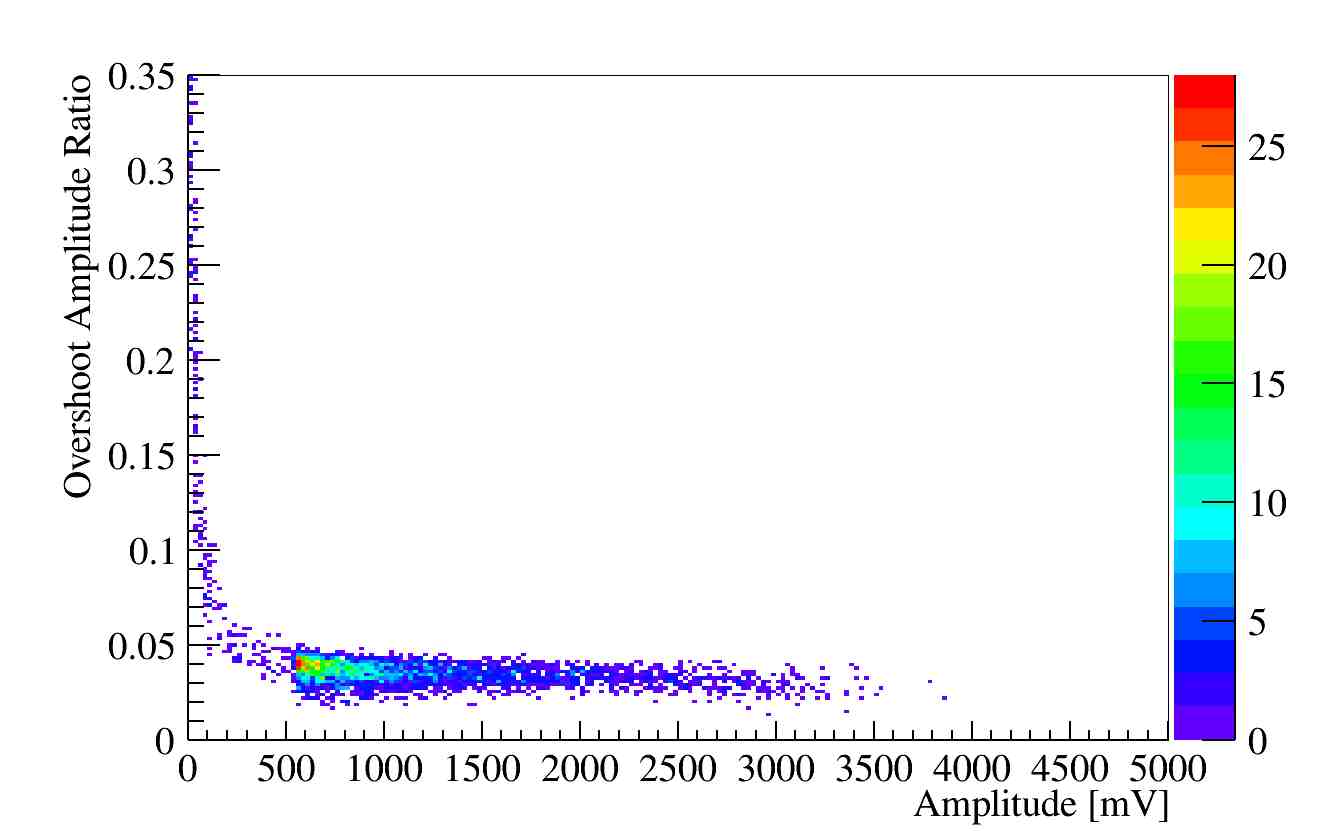}
	\caption{Dynode PMT}
	\label{fig:ampratio_ham}
	\end{subfigure}
    \caption{2-D distributions of the amplitude ratio of overshoot to primary pulse for one MCP PMT (a) and one dynode PMT (b) with coarse range.}
    \label{fig:ampratio}
\end{figure}

\begin{figure}[!htb]
    \centering
	\begin{subfigure}[c]{0.48\textwidth}
	\centering
	\includegraphics[width=\linewidth]{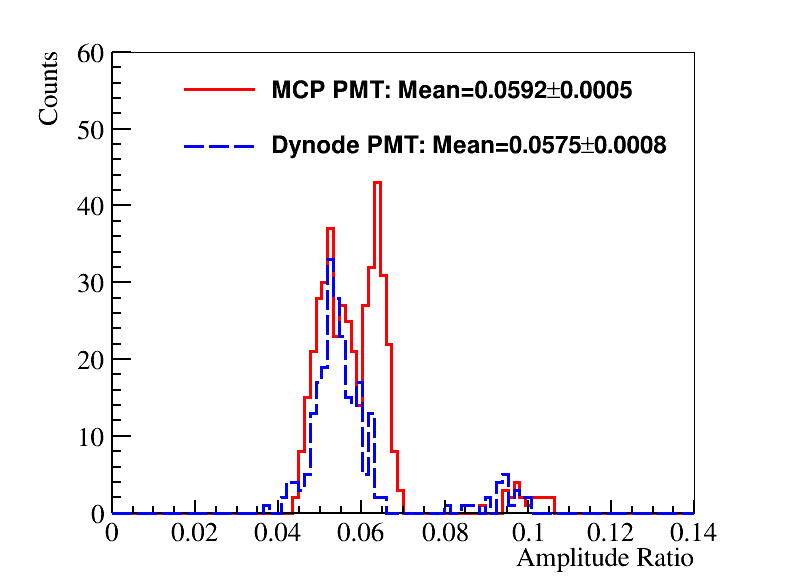}
	\caption{Fine range}
	\label{fig:overshoot:range1_ratio}
	\end{subfigure}
	\begin{subfigure}[c]{0.48\textwidth}
	\centering
	\includegraphics[width=\linewidth]{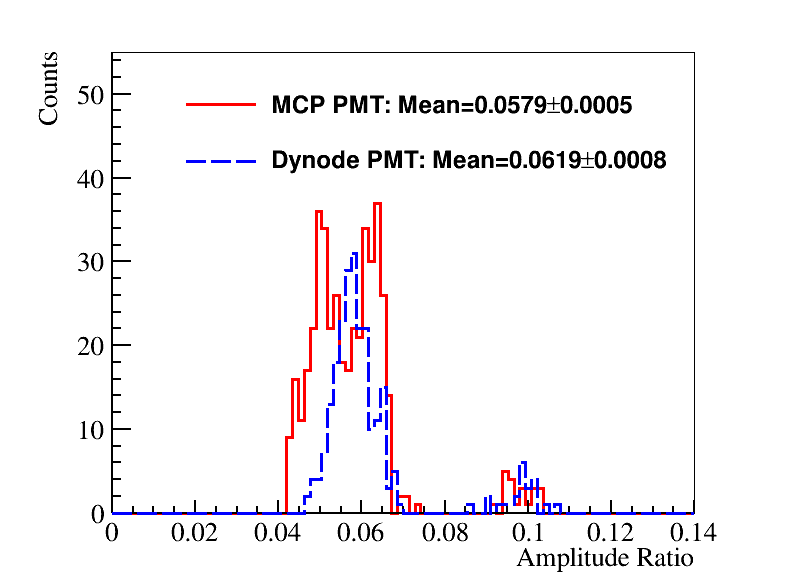}
	\caption{Coarse range}
	\label{fig:overshoot:range0_ratio}
	\end{subfigure}
    \caption{Overshoot ratio of MCP PMT (red histogram) and dynode PMT (blue histogram) in amplitude for the fine range (a) 
and the coarse range (b).}
    \label{fig:overshoot_ratio}
\end{figure}

Fig.\,\ref{fig:waveform} is a typical waveform from PMT self-trigger, where an overshoot can be observed at the end of the pulse. The overshoot arises from the discharging of the split capacitor for the HV and PMT pulse of the HV divider with positive HV\,\cite{JUNOPMTsignalopt}. Overshoot directly affects the linearity of PMT charge measurement, system triggering, and energy reconstruction. By optimizing the voltage divider, the amplitude of overshoot has been limited to around 1\%. However, in experimental engineering, due to the presence of extension cables, connectors, and impedance matching inside the electronics, the final amplitude of overshoot is expected to exceed this limit. Previous researches have indicated that the overshoot amplitude of waveform correlates with noise and signal amplitude\,\cite{Liu_2023}. The overshoot amplitude of small signals is susceptible to dark noise. In this study, we utilized data with a high threshold. Here, the overshoot amplitude is proportional to the signal amplitude. Fig.\,\ref{fig:ampratio} presents the variation of the overshoot amplitude ratio with signal amplitude for one MCP PMT and one dynode PMT with coarse range, analyzed without applying the electronic threshold. As the signal amplitude grows, the overshoot amplitude ratio remains relatively stable. Thus, obtaining the mean overshoot amplitude ratio for each PMT is sufficient. As shown in Fig.\,\ref{fig:overshoot_ratio}, the average overshoot amplitude ratio is 6\% in both the fine and coarse ranges. The small bulge at 9.5\% was found through waveform inspection to be due to the fact that the overshoot peak is significantly affected by reflection. Additionally, the results for the overshoot amplitude ratio for dynode PMT and MCP PMT are comparable.

\subsection{Combination of two dynamic range}
\label{2:consistency}

\begin{figure}[!htb]
    \centering
	\begin{subfigure}[c]{0.43\textwidth}
	\centering
	\includegraphics[width=\linewidth]{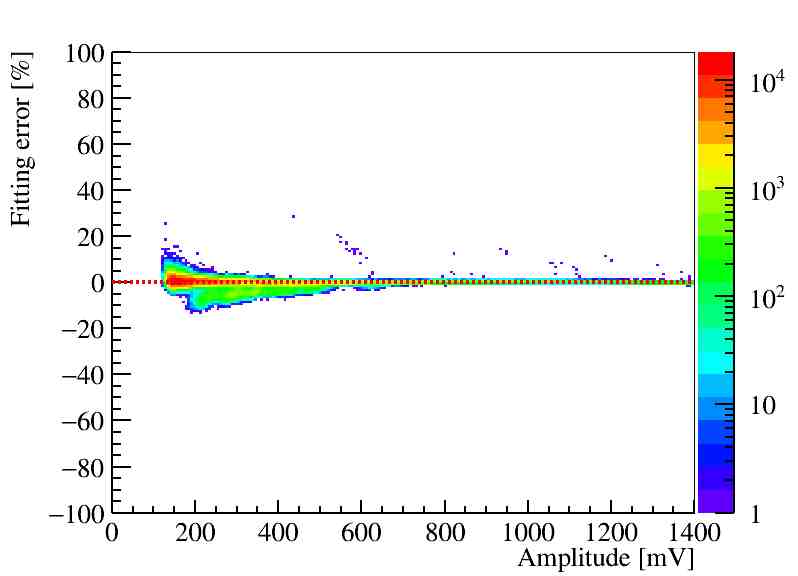}
	\caption{Amplitude}
	\label{fig:consistency:amp_fiterror}
	\end{subfigure}
	\begin{subfigure}[c]{0.43\textwidth}
	\centering
	\includegraphics[width=\linewidth]{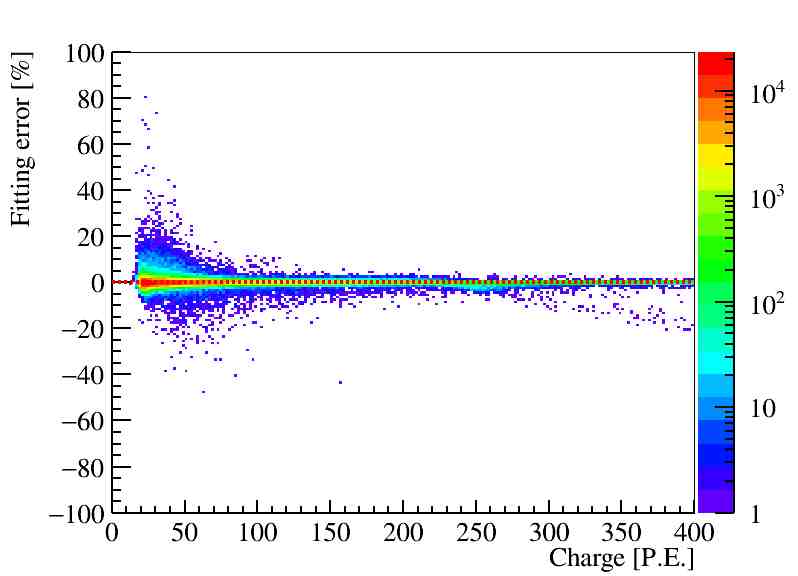}
	\caption{Charge}
	\label{fig:consistency:Q_fiterror}
	\end{subfigure}
    \caption{The relative fitting error of amplitude (a) and charge (b) between the fine range and coarse range.}
    \label{fig:consistency}
\end{figure}

\begin{figure}[!htb]
	\centering
	\includegraphics[width=0.6\linewidth]{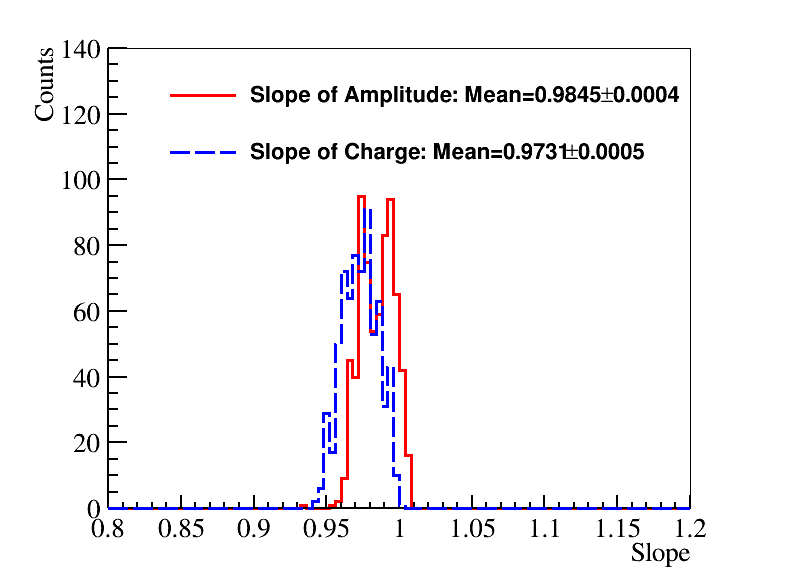}
    \caption{The consistency of amplitude (red histogram) and charge (blue histogram) between the two ranges.}
    \label{fig:consistency_slope}
\end{figure}

Since the waveforms output from the two ranges are actually the results of the same pulse but amplified at different gain factors, theoretically, the amplitude and charge of both should be consistent in the same unit. It was observed during the analysis that the upper measurement limits of the fine range and the coarse range are approximately 1380\,mV and 5000\,mV respectively. The waveforms exceeding these upper limits will exhibit saturation, which will lead to an underestimation of the charge value. Therefore, a first-order linear function is used to fit within the fine range, and the slope obtained from the fitting is used as the consistency parameter, as shown in Fig.\,\ref{fig:consistency_slope}. Within the fine range, both the amplitude and charge exhibit good consistency, with values of approximately 99\% and 97\%, respectively. In other words, the amplify factor of the two ranges is approximately 98\%. Using the same method, the fitting error for charge is larger than that for amplitude, as shown in Fig.\,\ref{fig:consistency}.

It is suggested to switch from fine range to coarse range when the amplitude of the pulse in fine range is higher than 1200\,mV or 1000\,ADCs.

\begin{figure}[!htb]
    \centering
    \includegraphics[width=0.6\linewidth]{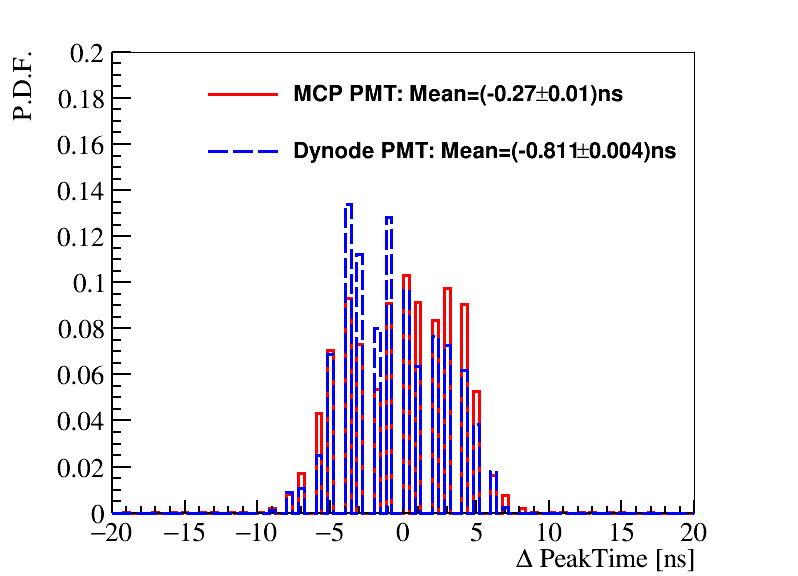}
    \caption{The difference of peak-time of MCP PMT (red histogram) and dynode PMT (blue histogram) between the two ranges.}
    \label{fig:detatime}
\end{figure}

The two ranges of the same channel are triggered by the same clock pulse. In theory, the hit-time of the two output pulses should be the same. However, due to the 8\,ns period of the clock pulse, the two ranges are not strictly triggered at the same point of the clock pulse. Therefore, as shown in Fig.\,\ref{fig:detatime}, there is an unpredictable and less than one clock pulse period difference in the hit-time of the pulses of the two ranges. This is independent of the type of PMT and is caused by the electronics itself.

\section{Summary}
\label{1:summary}

With the mass testing of JUNO 20-inch potted PMT with 1F3 electronics prototype at the Pan-Asia testing platform, we get a batch samples on the PMT self generated pulses. The rate is around 100\,Hz per PMT for both MCP or dynode PMTs on the ground with a threshold around 100\,mV (or 10\,P.E.). Besides, the time and charge features of the PMT self generated pulses are evaluated. Thanks to the broad amplitude distribution of the signals, we further investigated the time distribution of the after pulses, as well as the the variation of their number and charge with respect to the main pulse charge. Based on the data from fine range and coarse range, we examined the linearity between amplitude and charge, the timing of both dynamic ranges, and the consistency in amplitude and charge between the two ranges.



\section*{Acknowledgments}
\label{sec:1:acknow}
This work was supported by the National Natural Science Foundation of China No.\,11875282, the Strategic Priority Research Program of the Chinese Academy of Sciences, Grant No.\,XDA10011100, the CAS Center for Excellence in Particle Physics.

\bibliographystyle{unsrtnat}
\bibliography{main}   

\end{document}